\documentclass[12pt]{article}

\usepackage{epsf}

\newcommand{\bmat}{\left(\begin{array}}
\newcommand{\emat}{\end{array}\right)}
\def\NPB{Nucl. Phys. B}

\def\yzero{\smash{\hbox{$y\kern-4pt\raise1pt\hbox{${}^\circ$}$}}}

\def\beq{\begin{equation}}
\def\eeq{\end{equation}}
\def\beqa{\begin{eqnarray}}
\def\eeqa{\end{eqnarray}}

\def\-{\hphantom{-}}

\def\s2{\frac{1}{\sqrt2}}

\def\beq{\begin{equation}}
\def\eeq{\end{equation}}
\def\beqa{\begin{eqnarray}}
\def\eeqa{\end{eqnarray}}

\def\IF{\relax{\rm I\kern-.18em F}}
\def\II{\relax{\rm I\kern-.18em I}}
\def\IP{\relax{\rm I\kern-.18em P}}
\def\IC{\relax\hbox{\kern.25em$\inbar\kern-.3em{\rm C}$}}
\def\IR{\relax{\rm I\kern-.18em R}}

\def\cp{{\cal P}}

\def\Dsl{\,\raise.15ex\hbox{/}\mkern-13.5mu D} 
\def\IZ{Z\kern-.4em  Z}

 \def\cp#1{\relax\ifmmode {\IP\kern-2pt{}_{#1}}\else $\IP\kern-2pt{}_{#1}$\=fi}

\catcode`\@=11
\newdimen\@rotdimen
\newbox\@rotbox

\def\@vspec#1{\special{ps:#1}}
\def\@rotstart#1{\@vspec{gsave currentpoint currentpoint translate
   #1 neg exch neg exch translate}}
\def\@rotfinish{\@vspec{currentpoint grestore moveto}}
\def\@rotr#1{\@rotdimen=\ht#1\advance\@rotdimen by\dp#1%
   \hbox to\@rotdimen{\hskip\ht#1\vbox to\wd#1{\@rotstart{90 rotate}%
   \box#1\vss}\hss}\@rotfinish}
\def\@rotl#1{\@rotdimen=\ht#1\advance\@rotdimen by\dp#1%
   \hbox to\@rotdimen{\vbox to\wd#1{\vskip\wd#1\@rotstart{270 rotate}%
   \box#1\vss}\hss}\@rotfinish}%
\def\@rotu#1{\@rotdimen=\ht#1\advance\@rotdimen by\dp#1%
   \hbox to\wd#1{\hskip\wd#1\vbox to\@rotdimen{\vskip\@rotdimen
   \@rotstart{-1 dup scale}\box#1\vss}\hss}\@rotfinish}%
\def\@rotf#1{\hbox to\wd#1{\hskip\wd#1\@rotstart{-1 1 scale}%
   \box#1\hss}\@rotfinish}%
\def\rotate{\@ifnextchar[{\@rotate}{\@rotate[l]}}
\def\@rotate[#1]#2{\setbox\@rotbox=\hbox{#2}\@nameuse{@rot#1}\@rotbox}

\catcode`\@=12

\begin{document}

\makeatletter \@addtoreset{equation}{section} \makeatother
\renewcommand{\theequation}{\thesection.\arabic{equation}}
\pagestyle{empty}
\pagestyle{empty}
\rightline{FTUAM-03-09}
\rightline{IFT-UAM/CSIC-03-17}
\rightline{\today}
\vspace{0.1cm}
\setcounter{footnote}{0}

\begin{center}
\LARGE{
{\bf SU(5) Unified Theories
from 
Intersecting Branes}}
\\[2mm]
{\large{ Minos Axenides$^{1}$, Emmanuel Floratos$^{1,2}$ and Christos Kokorelis$^{3}$}}
\\[1mm]
 \normalsize{\em $^1$ Institute of Nuclear Physics, N.C.R.P.S. Demokritos,
GR-15310, Athens, Greece}\\
\normalsize{\em $^2$  Nuclear and Particle Physics Sector, Univ. of Athens,
GR-15771 Athens, Greece}\\
\normalsize{\em $^3$ Departamento de F\'\i sica Te\'orica C-XI and
Instituto de F\'\i sica
Te\'orica C-XVI},\\
{\em Universidad Aut\'onoma de Madrid, Cantoblanco, 28049, Madrid, Spain}
\end{center}
\begin{center}
{\small \bf Abstract}
\end{center}
We discuss the first string theory examples of 
three generation non-supersymmetric
 SU(5) and {\em flipped} SU(5) GUTS, which break to the Standard model
 at low energy,
 without extra matter and/or gauge group factors.
Our GUT examples are based on IIA $Z_3$ orientifolds with
D6-branes intersecting at non-trivial angles. These theories
necessarily satisfy RR tadpoles  and are free of NSNS tadpoles as
the complex structure moduli are frozen (even though a dilaton
tadpole remains) to discrete values. We identify appropriately the
bifundamental Higgses responsible for electroweak symmetry
breaking. In this way, the neutrino see-saw mechanism get nicely
realized in these constructions. Moreover, as baryon number is not a gauged
symmetry gauge mediated dimension six operators
do contribute to proton decay; however proton lifetime may be safely
enhanced by appropriately choosing a high GUT scale. An accompanying
natural doublet-triplet splitting guarantees the suppression of scalar
mediated proton decay modes and the stability of triplet scalar masses
against higher dimensional non-renormalizable operators.

\newpage

\setcounter{page}{1} \pagestyle{plain}
\renewcommand{\thefootnote}{\arabic{footnote}}
\setcounter{footnote}{0}

\section{Introduction}

Grand Unified Theories (GUTS) were historically formulated in
order to derive the Standard Model (SM), $SU(3) \times SU(2)
\times U(1)_Y$, from a single unified gauge group \cite{GG, pati1}
interaction \footnote{Embedding the SM into a  larger albeit
semi-simple gauge group was the original idea of \cite{pati1}}.
For more recent reviews see \cite{pati2, lan}. Novel phenomena
such rapid proton decay as well as the gauge hierarchy problem
motivated the development of N=1 supersymmetric (SUSY) unification
\cite{geodimo}. See also \cite{reports1} for reviews in the
development of supergravity GUTS that follow the early global SUSY
GUT era. The subsequent advent of superstring unification
\cite{green} motivated much work on embedding (SUSY) GUTS in the
framework of four dimensional compactifications of the heterotic
string \cite{total}. Important challenges that these
model building constructions still have to face are: abundance of extra matter and a large number
of undetermined scalar moduli, the sypersymmetry breaking
mechanisms etc.
However, apart from  providing a consistent framework for
perturbative quantum gravity, a standing goal of string theory has
been to demonstrate, that vacua with only the SM at low energy
exist. Steps towards bypassing  this obstacle have been taken
recently through the use of D-brane machinery \cite{pol} , in
model building constructions based on intersecting brane worlds
(IBW's) \cite{kokos1}-\cite{nano}. There, in a top-bottom
approach, it
 became possible to localize, at four dimensions, the Standard Model
with right handed neutrinos \cite{louis2, kokos5, kokos6}. We note
that models with only the SM at low energy were also derived from
a rather more arbitrary non-string bottom-up approach \cite{toma}.
In the context of IBW's, the first string theory examples of four
dimensional (4D) three generation (3G) non-SUSY string GUT models
which break to exactly the Standard model at low energies without
extra matter and/or gauge group factors were constructed in
\cite{kokos1}. The most important phenomenological features of
these classes of models \footnote{, which are based on the
Pati-Salam gauge group structure $SU(4)_C \times SU(2)_L \times
SU(2)_R$,}, which have D6-branes intersecting at angles on a
toroidal orientifold \cite{lust1}, may be briefly described as
follows :
\newline $\bullet$ The models are built on an orientifolded toroidal
background \cite{lust1}, with the a; b branes (supporting a $U(4)_C$; $U(2)_L$ gauge group resp.) accommodating
the quarks, leptons; antiquarks respectively. Also some `basic' branes
$c, d, \cdots$ are needed to be
added, whose fermionic content is enforced by the RR tadpole
\footnote{Model based on D6-branes intersecting at angles  are T-dual to
models with magnetic
deformations \cite{ang, pra}.} cancellation
conditions. Their solution, in turn, enforces the introduction of
extra $e^i$ branes.
\newline
$\bullet$ Baryon number is a gauged symmetry and thus
proton is stable. The theory has overall $N=0$ supersymmetry (SUSY) in
its open sectors, but it allows the existence of N=1 SUSY
in some open string sectors. The latter sectors may respect N=1 SUSY in
order, e.g. to allow
for the existence of a Majorana mass for the right handed neutrino.
Also N=1 SUSY is required to make massive, by creating gauge singlets,
the extra fermions appearing
from the non-zero intersections between the extra branes $e^i$
and the branes $c, d, \cdots$.
\newline
$\bullet$ The most important result of these non-supersymmetric
GUT constructions is  the existence of N=1 SUSY in some open
sectors which gives rise to relations among  the parameters of the
RR tadpole cancellation conditions. The latter simultaneously
solve the conditions of the extra U(1)'s (beyond the SM one's)
which survive massless the Green-Schwarz anomaly cancellation
mechanism.
We note that models based on toroidal orientifold GUTS
\cite{kokos1}, even though they become automatically free of
tachyons have as an open question the issue of their full
stability as NSNS tadpoles are being left over \cite{lust3}.
Nevertheless, as no definite statement can be made about the full
stability of non-supersymmetric toroidal orientifolds, in the present work 
we turn out attention to non-supersymmetric 4D chiral
SU(5) and flipped SU(5) GUT orientifold constructions  which involve
$D6$ branes intersecting at angles and possess orbifold
symmetries \cite{lust3}. In this way, complex structure moduli
will be fixed from the orbifold symmetry and the associated
tadpoles vanish. We note that N=1 supersymmetric SU(5) models have been first 
constructed in \cite{cve0} and systematically explored in \cite{cve1}, 
even though at present there are no  
models which are free of exotic massless matter, present, to low energies. 
On the other hand non-SUSY SU(5) GUTS have been considered 
before on \cite{lust3}.
These models were suspected to suffer from serious shortcomings
such as the absence of an appropriate identification for the
electroweak (EW) 5-plets. Indeed both the tree level $SU(5)$
couplings $ {\bf 10} \cdot {\bf 10} \cdot {\bf 5}$
 responsible for giving masses
to the $(u, c, t)$ quarks and the Dirac mass terms needed for the
realization of the see-saw mechanism were not allowed
\footnote{The absence of the mass couplings $ {\bf 10} \cdot {\bf
10} \cdot {\bf 5}$ persists in the 4D N=1 supersymmetric $U(5) \times
U(1)^n$ standard-like models of \cite{cve1}.}. The flipped SU(5)
GUTS were also considered previously \cite{nano} in IBW's.
The proposed models, however, have several problems including the
presence of extra massless matter at low energies \cite{nano}.
In this work, we take a fresh look in the construction of SU(5)
and flipped SU(5) GUTS. In particular, we will focus our attention
to the construction of phenomenologically interesting SU(5)
\cite{GG} and flipped SU(5) \cite{bar,AA} non-supersymmetric GUTS,
that break to exactly the Standard Model (SM) at low energies
without extra matter and/or additional gauge group factors
\footnote{We note that even these models are overall non-SUSY they
may have in cases N=1 supersymmetric sectors that may offer a
partial protection of the Higgs doublets against the quadratic
mass renormalization.}. The models we will present do not have
baryon number as a gauged symmetry. A resolution to the
doublet-triplet splitting problem thus guarantees
phenomenologically acceptable long proton lifetime. We will
identify appropriately the EW pentaplet Higgses for the presented
classes of GUTS \cite{lust3} and demonstrate the presence of a
see-saw mechanism. To that end, we show that the existence of
electroweak ${5}$-plets is independent of the existence or not of
tachyon bifundamental scalars. The latter fields break, in the
SU(5) GUTS, the extra U(1) which survive massless the
Green-Schwarz anomaly cancellation mechanism.
The paper is organized as follows:
 In section two
 we describe the general rules for
 building chiral GUT models in four dimensional $Z_3$ orientifolded $T^6$
 compactifications of IIA theory with D6 branes intersecting at angles.
 In section 3, we discuss the
 building of the usual SU(5) GUTS
 in these constructions by
 identifying appropriately the 5-plet electroweak
 Higgses. We also discuss the realization of the see-saw mechanism as well as 
the presence of
Yukawa couplings in the above models.
The second part of this work, is focused on the
construction of flipped SU(5) GUTS from intersecting branes.
In section 4, we describe explicitly the construction, fermion spectra,
 of the new classes of
flipped
 SU(5) GUTS which break to the low energy without
gauge group factors and/or extra fermions.
 In subsection 4.1, we discuss the GUT set of Higgses in the
models and in subsection 4.2 the specific realization of the electroweak Higgs
as previously massive tachyon bidoublets. In section 5, we discuss
the issue of quark-neutrino-lepton masses in flipped SU(5) GUTS. In section 6 we discuss
issues related to the proton decay modes in general SU(5)
constructions from intersecting brane worlds, where baryon numbers is not
a gauged symmetry. We discuss first the SU(5) GUTS. Then the flipped SU(5)
GUTS are discussed,
where the mechanism of doublet-triplet splitting mechanism,
in a non-supersysmmetric
framework,
which suppress scalar mediated proton decay modes, is
tested against its stability due to higher order non-renormalizable interactions.
 Finally in section 7, we present our conclusions.

\section{Geometry of $\frac{T^6}{\Omega \times {\cal R} \times Z_3}$
orientifolds in intersecting brane worlds}

We consider type I theory with D9 branes with magnetic fluxes.
Also let us assume that we have a six dimensional tori, which is
decomposed 
as a product of three two dimensional tori
$T^6 = (T^2)^I \otimes (T^2)^J \otimes(T^2)^K$ and where we have
introduced
complex
coordinates $Z^l = X^l + i Y^l$, $l = I, J, K$ defined on the three tori.
After performing a T-duality along e.g. the $Y^l$ direction our theory
becomes an orientifold of type IIA with D6-branes intersecting at
non-trivial angles. Imposing a further left-right symmetric $Z_3$ orbifold
symmetry $\theta : \ Z^l \rightarrow e^{\frac{2\pi i}{3}}Z^l$, our
theory becomes an orientifold of
\beq
\frac{(Type \ IIA) /Z_6}{\{Z_3 + \Omega R Z_3\} } \ .
\label{ena1}
\eeq
In order to cancel the orientifold charge 
in six dimensional lattices, which have \footnote{The A lattice is
defined by the complex structure moduli
$U^A = \frac{1}{2}+ i
\frac{\sqrt{3}}{2} $,
while the complex structure of the
B-lattice is defined by $U^B = \frac{1}{2}+ i
\frac{1}{2 \sqrt{3} } $. See \cite{lust3} for more details.}
the form
AAA AAB, ABB, BBB,
we introduce D6-branes at arbitrary angles which satisfy 
the following RR tadpole conditions \cite{lust3}
  \beq
\sum_a N_a Z_a = 2
\label{tad}
 \eeq
The net number of bifundamental massless chiral fermions in the models
is defined as
\beqa
({\bar N}_a, N_b)_L :\  I_{ab} = Z_a Y_b - Y_a Z_b \\
(N_a, N_b)_L : \   I_{ab^{\star}} = Z_a Y_b + Y_a Z_b
\label{spec1}
\eeqa
Also present in the models are chiral fermions transforming
in the symmetric (S) and antisymmetric (A) representations of the gauge
group as follows
\beqa
(A + S)_L &=& Y_a (Z_a  - \frac{1}{2})\\
(A)_L &=& Y_a
\label{spec2}
\eeqa
These S, R representations appear after the
localization of open strings stretching between the $D6_a$
brane and its $\Omega {\cal R} \Theta^k$ images. Also non-chiral matter
is generally present,
but since is non-chiral it will be of no interest to us.
In general it becomes massive by pairing up with adjoint scalars.
The $Z_a, Y_a$, are characterized as effective wrapping numbers,
which are functions of the wrappings $(n^i, m^i)$, $i=1,2,3$; define
the homology cycle that the D6 brane wraps around the six dimensional
toroidal internal space.

Also due to a generalized Green-Schwarz mechanism that involve $BF$ type of
couplings, the combination
of U(1) $F_a$ fields
\beq
\sum_a N_a Y_a F_a
\label{green}
\eeq
gets massive by having a non-zero coupling to the corresponding RR field.

\section{$SU(5)$ GUTS from intersecting branes }

In \cite{lust3} the simplest construction of an SU(5) GUT was
partially considered. In its minimal version the construction
involves two stacks of D6-branes at the string scale $M_s$, the
first one corresponding to a $U(5)$ gauge group while the second one to a
U(1) gauge group. Its effective wrapping numbers are \beq (Y_a,
Z_a) = (3, \frac{1}{2}), \  (Y_b, Z_b) = (3, -\frac{1}{2}),
\label{ena} \eeq Under the decomposition $ U(5) \subset SU(5)
\times U(1)_a$, the models become effectively an $SU(5) \times
U(1)_a \otimes U(1)_b$ class of models. One combination of U(1)'s
become massive due to its coupling to the corresponding RR field. 
Another one remains massless to low energies. The latter
combination will be finally broken by an appropriate vev as we
will comment later.

The spectrum of these models is given in Table (\ref{spec3}),
\begin{table}[htb]\footnotesize
\renewcommand{\arraystretch}{1.5}
\begin{center}
\begin{tabular}{||c|c|c|c|c|c||}
\hline \hline
Sector name & Multiplicity & $SU(5)$    &  $U(1)_a$ & $U(1)_b$
& $U(1)^{mass}$ \\\hline
\{ 51 \} & $3$          & ${\bf {\bar 5}}$ & $-1$      &  $1$   & $-\frac{6}{5}$     \\\hline
$A_{a}$ & $3$          & ${\bf 10}$   &  2         &   0      &$\frac{2}{5}$          \\\hline
$S_{b}$ & $3$          &  ${\bf 1} $         &   0        &   -2   & $2$    \\\hline
  \hline
\end{tabular}
\end{center}
\caption{Chiral Spectrum of a two D6-brane stack three generation
$SU(5) \otimes U(1)^{mass}$ with right handed neutrinos model.
Also shown: the charges under the $U(1)^{mass(less)}$ gauge symmetry, which
 when rescaled
appropriately (and remains unbroken) converts the original $SU(5)$ model to the three
generation flipped $SU(5) \times U(1)^{fl}$ of table (\ref{flip}).
\label{spec3}}
\end{table}
where the first two rows have the chiral content of the standard
SU(5) generations and the third row is associated with the
presence of right handed neutrino. Choices of values for the
$(n^i, m^i)$ wrappings satisfying the RR tadpoles (\ref{tad})
include e.g. for the AAA lattice \beq U(5): \ [(-3, 2)(0, 1)(0,
-1)]; \  \ U(1): \ [(-3, 2)(1, -1)(-1, 0)] \label{wrap1} \eeq or
 \beq U(5): \ [(0, 1)\ (-1, 1)\ (1, -3)]; \ \ U(1): \ [(-1, 1)\
(1, 0)\ (3, 1)] \label{wrap2} \eeq The anomaly free U(1) symmetry
surviving massless the Green-Schwarz mechanism (see (\ref{green}))
is given by \beq U(1)^{mass} =\frac{1}{5}U(1)_a - U(1)_b \ .
\label{free} \eeq The latter symmetry may be broken when the
massive scalar superpartner of the right handed neutrino becomes
massless and develops a vev. As was mentioned in \cite{lust3} that
may always happen as a tachyonic direction $s^B = {\bf
1}^B_{(0,-2)}$ may always appear by open strings stretching
between the branes that support the orbit of $U(1)_b$. In this
case, the extra U(1) (\ref{free}) gets broken by the $\langle s^B
\rangle$ and the $U(5) \times U(1)$ model becomes an SU(5) class
of GUTS with right handed neutrinos. In this case, the low energy chiral
matter multiplet content, of table (1), is as follows : \beq {\bf
10} = (Q, u^c, e^c), \ {\bf {\bar 5}} = (d^c, L); \ {\bf 1} =
{\nu^c} \label{GUT12} \eeq

\subsection{ GUT and Electroweak Higgs Sector}

After the breaking of the  $U(1)^{mass}$ gauge symmetry, we
still face the following two problems. Firstly we have to break the SU(5)
symmetry down to the SM gauge symmetry and secondly to find the Higgs
${\bf 5}$-plets needed for electroweak symmetry breaking.\newline
The first problem is solved by the use of the adjoint scalar {\bf 24},
part of the N=4 Yang-Mills in the aa-sector, which breaks the GUT symmetry to the
$SU(3) \times SU(2) \times U(1)_Y$ \cite{lust3}. \newline
On the other hand, the simultaneous presence of electroweak 5-plets was
excluding the simultaneous presence of tachyon $1^B$ singlets,
and vice-versa, in all lattices used in \cite{lust3}.\newline
Lets us now see how the latter problem may be evaded in a way independent of the
particular lattice involved.
The electroweak ${\bf 5}$-plets needed, may be found
in the NS sector $\{ 51^{\star} \}$. We remind the alert reader
that there are no chiral fermions
contributing to the spectrum of the models from
the R sector $\{ 51^{\star} \}$.
Hence the EW set of Higgs may be found as part of the
massive N=2 hypermultiplet spectrum localized in this sector.
These bifundamental Higgs states may be localized as they become tachyonic,
as we vary the distances between the parallel $a$, $b$ branes.
They may be
seen \footnote{We note that identification of electroweak Higgses may
proceed
along the same lines in the N=1 supersymmetric SU(5) models of \cite{cve1}.}
in table (\ref{Higgs}).
\begin{table} [htb] \footnotesize
\renewcommand{\arraystretch}{1}
\begin{center}
\begin{tabular}{||c|c|c|c|c|c||}
\hline
\hline
Intersection & EW Higgs & repr. & $Q_a$ & $Q_b$ & $U(1)^{mass}$\\
\hline\hline
$\{ 51^{\star} \}$ & $h_1$  &  ${\bf 5}$   & $1$ & $1$ & $ -\frac{4}{5}$\\
\hline
$\{ 51^{\star} \}$  & $h_2$  &  ${\bf {\bar 5}}$   & $-1$ & $-1$ & $\frac{4}{5} $\\
\hline
\hline
\end{tabular}
\end{center}
\caption{\small Higgs fields responsible for EW symmetry breaking in
the SU(5) GUT.
\label{Higgs}}
\end{table}
We note that, for two branes belonging to different homology classes
$[5]$, $[1]$, associated with the $U(5)_a$, $U(1)_b$ D6-branes
respectively, the ({\em mass})$^2$ operator receives contributions from
open strings stretching between the a-brane and the
$\Omega {\cal R} \Theta^k$ image
of the b-brane. Thus the electroweak ${\bf 5}$-plets will receive
contributions from the three different sectors just described.

The typical form of such a contribution to the ({\em mass})$^2$
operator in the $\{ 51^{\star} \}$ sector is depicted in table (\ref{mass1}).
\begin{table} [htb] \footnotesize
\renewcommand{\arraystretch}{1}
\begin{center}
\begin{tabular}{||c|c|c||}
\hline
\hline
Intersection & State & $(Mass)^2$ \\
\hline\hline
  $\{ 51^{\star} \}$   &   $ h^{-} \rightarrow (-1+\vartheta_1, 0, \vartheta_3, 0)              $ &  $     \alpha^{\prime} {\rm (Mass)}_{h^{-}}^2 =
  \frac{Z_2^{(bc)}}{4\pi^2}\ +\
  \frac{1}{2}(\vartheta_3 - \vartheta_1)                  $                \\
\hline
  $\{ 51^{\star} \}$          & $ h^{+} \rightarrow (\vartheta_1, 0, -1+\vartheta_3, 0) $  
&   $\alpha^{\prime} {\rm (Mass)}_{h^{-}}^2 =
  \frac{Z_2^{(bc)}}{4\pi^2}\ +\
  \frac{1}{2}(\vartheta_1 - \vartheta_3)  $                          \\
\hline
\hline
\end{tabular}
\end{center}
\caption{\small States and $(Mass)^2$ in the SU(5) GUTS from intersecting 
branes
\label{mass1}}
\end{table}
Therein $Z_2^{(bc)}$ is the relative distance in
transverse space along the second torus from the orientifold plane;
$\vartheta_1$, $\vartheta_3$, are the (relative) angles
between the $5_a$, $1_{b^{\star}}$ D6 branes in the
first and third complex planes.
Note that we have made the assumption that the $5_a$, $1_{b^{\star}}$ D6 branes are parallel across
the second complex plane. In this way it will become transparent in what 
follows which is the effective Higgs combination in this particular orbit.
In fact, one may show that by using a particular set of wrappings, e.g.
the one's in (\ref{wrap1}) that the ({\em mass})$^2$ operator receives
contributions from the three different image orbits $\Omega {\cal R} \Theta^k$,
and that the
pair of `rotated' $5_a$, $1_{b^{\star}}$ D6 branes are parallel
along different complex planes,
across the different image orbits.

The presence of scalar doublets $h^{\pm}$
defined as
\beq
h^{\pm}={1\over2}(h_1^*\pm h_2) \   \ .
\label{obv1}
\eeq
can be seen
as
coming from the field theory mass matrix
\beq
(h_1^* \ h_2)
\left(
\bf {M^2}
\right)
\left(
\begin{array}{c}
h_1 \\ h_2^*
\end{array}
\right) + h.c.
\eeq
where
\beqa
{\bf M^2}=M_s^2
\left(
\begin{array}{cc}
Z_{2}^{(bc)}(4\pi^2)^{-1}&
\frac{1}{2}|\vartheta_1^{(bc)}-\vartheta_3^{(bc)}|  \\
\frac{1}{2}|\vartheta_1^{(bc)}-\vartheta_3^{(bc)}| &
Z_{2}^{(bc)}(4\pi^2)^{-1}\\
\end{array}
\right),
\eeqa
Hence the effective potential which
corresponds to the spectrum of electroweak
Higgs $h_1$, $h_2$ may be expressed as
\beqa
V_{Higgs}^{bc}\ =\ \overline{m}_H^2 (|h_1|^2+|h_2|^2)\ +\
(\overline{m}_B^2 h_1 h_2\ +\ h.c)
\label{bcstarpote}
\eeqa
where
\beqa
\overline{m}_H^2 \ =\ \frac{Z_2^{(bc)}}{4\pi^2\alpha'} \
& ;&
\overline{m}_B^2\ =\ \frac{1}{2\alpha'}|
\vartheta_1^{(bc)} - \vartheta_3^{(bc)}|
\label{bchiggs}
\eeqa
It is obvious from (\ref{obv1}) that the presence
of EW ${\bf 5}$-plets of table (2), may be interpreted from the low energy
point of view as an effective combination of the
fields $h_1$ with charges $(1, 1)$ and its
conjugate
representation $h_2$ with charges  $(-1, -1)$. The latter comment
will be particularly useful in the construction of flipped SU(5) GUTS
later on.

\subsection{ Mass generation and Yukawa couplings}

In this section, we will discuss the allowed  Yukawa couplings
responsible for giving masses to the up-down quarks. We also
discuss the realization
of the see-saw mechanism. The existence of this mechanism is
particularly important for the phenomenology of the models as it
may be responsible at disk level, for generating small neutrino
masses in consistency with the neutrino oscillation experiments.

\begin{itemize}
\item { See-Saw mechanism}
\end{itemize}
In the SU(5) theories described by the chiral spectrum of table (1),
baryon and lepton number is not a gauged
symmetry but rather it is the combination {\em (B-L)} which becomes gauged.
This means that physical processes violate baryon and lepton number, 
e.g. $\triangle(B-L) = 0$. The violation of lepton number (VLN), e.g.
by two units, may be seen as
one of the most difficult
problems in model building, namely the generation of small neutrino
masses. The VLM may be easily accommodated in the presence
of the Majorana mass term for right handed neutrinos in the 
see-saw mechanism.

In the presence of a lepton number violating term describing
the right handed neutrino the Yukawa interactions give rise to the see-saw
mechanism. Within our choice of electroweak Higgses takes
 the usual form \
\beqa
&
{\cal L} = Y^{\nu_L \nu_R}_{ij} \cdot  {\bar  {\bf 5}}_i \cdot {\bf 1}_j \cdot
{\bf { 5}}^B \
      + \  Y^{\nu_R} \cdot \frac{1}{M_s} \cdot {\bf 1}_i \cdot {\bf 1}_j \cdot {\bf {\bar 1}}_i^B \cdot {\bf {\bar 1}}_j^B\nonumber\\
or&
{\cal L} = Y^{\nu_L \nu_R}_{ij} \cdot {\bar {\bf 5}}_i \cdot 
{\bf 1}_j \cdot  
{ h}_2 \
      + \  Y^{\nu_R} \cdot \frac{1}{M_s} \cdot {\bf 1}_i \cdot {\bf 1}_j \cdot {\bf {\bar 1}}_i^B \cdot {\bf {\bar 1}}_j^B
\label{seesaw}
\eeqa

The standard version of the see-saw mass matrix \beq \left(
\begin{array}{c}
\nu \\ \nu^c
\end{array}
\right)
\left(
\begin{array}{cc}
0 & m\\
m& M
\end{array}
\right)
\left(
\begin{array}{c}
\nu \ \nu^c
\end{array}
\right) \label{see} \eeq is realized with \beq m = Y^{\nu_L \nu_R}
\cdot \ \langle { h}_2 \rangle \sim Y^{\nu_L \nu_R}\cdot
\upsilon \ , \ M = Y^{\nu_R} \cdot \frac{\langle {\bf {\bar
1}}_j^B \rangle \cdot \langle {\bf {\bar 1}}_j^B \rangle}{M_s}
\sim \ M_s, \label{inter} \eeq where the indices $i, j$ are
generation indices and we have chosen $Y^{\nu_L \nu_R} \  =
e^{-A_1}$, $ Y^{ \nu_R} \sim e^{-A_2}$ and the hierarchy of
neutrino masses is generated 
\footnote{As it has been shown explicitly in
the first reference of \cite{kokos1}, different choices of
wrapping numbers involved in the leading worldsheet area contributions
to the see-saw mass matrix Yukawa couplings, may always
generate a hierarchy of neutrino masses in consistency with
neutrino oscillation experiments.}
by an interplay of the areas involved
in in (\ref{inter}). The eigenvalues of the see-saw
(\ref{see}) are the heavy one $M_1$, which corresponds to the right handed
neutrino, and the light one, $M_2$ which corresponds to the left
handed neutrino \beq
 M_1 \ = \ M,  \ \ \ \ M_2  =\  \frac{m^2}{M_1} \ .
 \label{corre}
  \eeq

The presence of the see-saw mechanism, renders the
right handed neutrino sufficiently massive (of the order of the 
string scale $M_s$),
such that it cannot be found at scales of the electroweak order.
Therefore after the implementation of the see-saw (\ref{see}) and the
breaking of the U(1) (\ref{free}),  one is left with an SU(5) type of
GUT,
 which breaks with the use of the 
adjoint vector multiplet, a $\bf 24$, to the 
$SU(3) \times SU(2) \times U(1)_Y$. 
Subsequently, the EW bifundamental Higgs $\bf 5$-plets of table
(\ref{Higgs}) break the electroweak symmetry.



$\bullet$ { Masses for down quarks}


Lets us focus our attention first to the d- type quarks. The
Yukawa couplings responsible for giving masses to the d-type
quarks \footnote{we are interested only in the couplings of the
first generation.} are given by the `tree' level expression \beq Y^d \cdot {\bf 10} \cdot {\bf
{\bar 5}}  \cdot {\bf {\bar 5}}^B = \ Y^d  \cdot \langle h_2
\rangle \cdot {\bf 10} \cdot {\bf {\bar 5}} \label{asume1} \eeq
The factor $Y^{d} $ parametrizes \footnote{$Y^{u}$ in the case of
u-type Yukawa's. } the classical dependence of the Yukawa
couplings on the worldsheet area A connecting the three
vertices. As this coupling is of order $e^{-A} \cdot
\upsilon$, the physical mass of the d-quark, $m_d = 0.01$ GeV is
reached naturally from above by choosing the area $A= 10.11$.

$\bullet$ { Masses for up quarks}


The candidate `tree' level coupling for giving mass to the
$(u, c, t)$ quarks given by
 ${\bf 10} \cdot
{\bf 10} \cdot {\bf 5}$ does not exist from charge conservation. 
The same coupling is also absent in attempts to build a
consistent 3G 4D SU(5) model in \cite{lust3} and also
 in 3G 4D supersymmetric SU(5)-like constructions \cite{cve1}. 
Necessarily,
the mass term responsible for up-quark generation may come from higher
order non-renormalizable couplings.

\section{Construction of {\em flipped} SU(5) GUTS}

In \cite{nano} it was noticed that if one leaves unbroken
the $U(1)^{mass}$ factor, which remains 
massless after the implementation of the Green-Schwarz 
anomaly cencellation mechanism, and its charge is appropriately rescaled, 
the chiral content of table (1) is that of a three generation flipped
SU(5) model. The relevant charges for the flipped SU(5) model may be seen
in the last column of table (\ref{flip}).
\begin{table}[htb]\footnotesize
\renewcommand{\arraystretch}{1.5}
\begin{center}
\begin{tabular}{||c|c|c|c|c|c|c|c||}
\hline \hline
Field & Sector name & Multiplicity & $SU(5)$    &  $U(1)_a$ & $U(1)_b$
 & $U(1)^{mass}$ & $U(1)^{fl} = \frac{5}{2} \times U(1)^{mass} $\\\hline
$f$& \{ 51 \} & $3$          & ${\bf {\bar 5}}$ & $-1$      &  $1$   & $-\frac{6}{5}$ & -3    \\\hline
$F$ & $A_{a}$ & $3$          & ${\bf 10}$   &  2         &   0      &$\frac{2}{5}$ &   1          \\\hline
$l^c$ & $S_{b}$ & $3$          &  ${\bf 1} $         &   0        &   -2   & $2$ & 5    \\\hline
  \hline
\end{tabular}
\end{center}
\caption{Chiral Spectrum of a two intersecting D6-brane stacks in a three 
generation flipped $SU(5) \otimes U(1)^{mass}$ model.
Note that the charges under the $U(1)^{fl}$ gauge symmetry,
 when rescaled
appropriately (and $U(1)^{fl}$ gets broken) `converts' the flipped SU(5) model
to the three
generation $SU(5)$ of table (\ref{spec3}). 
\label{flip}}
\end{table}
However, the 3G flipped SU(5) models produced in \cite{nano}
have extra massless chiral matter remaining at low energies
scales of the order of the electroweak scale
and an incomplete GUT and/or electroweak (EW) Higgs sector.

In this section, we will achieve the task of obtaining a 3G flipped SU(5)
GUT from intersecting D6 branes without extra
matter and with a complete GUT and EW Higgs sector. Thus these models will
achieve naturally the breaking to the SM gauge group of electroweak
interactions, without extra matter and/or gauge group factors. Subsequently the electroweak Higgses will break the SM to the $SU(3)_c \times U(1)_{EM}$.

In a SU(5) GUT with the flipped chiral content, the fifteen fermions of the SM plus the right
handed neutrino $\nu^c$ belong to the \cite{bar} 
\beq
F = {\bf 10_1} = (u, d, d^c, \nu^c), \ \
f = {\bf {\bar 5}_{-3}} = (u^c, \nu, e), \  \ l^c =  {\bf 1_{5}} = e^c
\label{def1}
\eeq
chiral multiplets.
On the other hand the flipped SU(5) gauge group needs to be broken at the grand unification
scale $M_{GUT}$ by the vacuum expectation values of two fields \footnote{The
subscript denotes the U(1) charge under the $U(1)^{fl}$ factor of table
(\ref{flip}).} ${\bf 10_1^B}$, ${\bf 10_{-1}^B}$, where
\beq
H = {\bf 10_1^B} = (u_H, d_H, d^c_H, \nu^c_H ),\
\ {\bar H} = {\bf \overline{10}_{-1}^B} = ( {\bar u}_H, {\bar d}_H,
{\bar d}^c_{ H}, {\bar \nu}^c_{ H}  ) \ .
\label{def2}
\eeq
 Also, the electroweak
symmetry breaking needs a Higgs sector that guarantees the presence
of the multiplets ${\bf 5_2^B}$, ${\bf {\bar 5}_{-2}^B}$
\beq
h_3 = {\bf 5_{-2}^B} = (D_1, D_2, D_3, h^{-},  h^{0} ),\
\ h_4 = {\bf {\bar 5}_{2}^B} = ( {\bar D}_1, {\bar D}_2,
{\bar D}_3, h^{+},  {\bar h}^{0})  \ .
\label{electro}
\eeq
In what follows, we demonstrate that these
two issues will be
dealt with and solved independently of the presence of lattices 
on which the orbifold symmetry acts.
The U(1) charge of the flipped  $SU(5) \times U(1)$ GUT from
intersecting D6-branes is depicted in table (\ref{flip}).

\subsection{ {GUT Higgses for flipped SU(5)}}

GUT breaking Higgses may be found in the `vicinity' of the chiral spectrum
of table (\ref{flip}).
The necessary Higgs GUT multiplets may come from the massive spectrum
of the sector localizing
the ${\bf 10}$-plet fermions seen in table (\ref{flip}). The lowest order
Higgs in this sector, let us call them $H_1$, $H_2$ have quantum numbers as
those given in table (\ref{Higgsfli}).

\begin{table} [htb] \footnotesize
\renewcommand{\arraystretch}{1}
\begin{center}
\begin{tabular}{||c|c||c|c|c|c||}
\hline
\hline
Intersection & GUT Higgses & repr. & $Q_a$ & $Q_b$ & $Q^{fl}$\\
\hline\hline
$\{ a,{\tilde O6} \}$  & $H_1$  &  {\bf 10}   & $2$   & $0$ & $1$ \\
\hline
$ \{ a,{\tilde O6} \}$  & $H_2$  &  ${\bf {\bar 10}}$   & $-2$ & $0$  & $-1$ \\
\hline
\hline
\end{tabular}
\end{center}
\caption{\small 
Flipped $SU(5) \otimes U^{fl}$ GUT symmetry breaking
scalars which are part of the massive excitation spectrum in the
appropriate sector. The last column denotes the flipped U(1) charge; the latter
is defined explicitly in table (\ref{flip}).
\label{Higgsfli}}
\end{table}
By looking at the last column of table (\ref{Higgsfli}), we realize that the
Higgs $H_1$, $H_2$ are the GUT symmetry breaking Higgses of a standard
flipped SU(5) GUT.
By dublicating the analysis of section (3.1),
one may conclude that what it appears in the effective theory as GUT 
breaking Higgs scalars, is the linear
combination
\beq
H^G = H_1 + H_2^{\star}  \ .
\label{fliGUT}
\eeq
Hence, if for example the number of GUT Higgses is found to be $n_H$,
that means that we have $n_H$ intersections, with $H^G$ scalars
localized at each intersection.

 Thus the necessary GUT Higgses for the breaking of the flipped
$SU(5) \otimes U(1)^{fl}$ get localized \footnote{
In \cite{nano} extra sectors, localizing chiral fermions in the
$10_1$, ${10_{-1}}$ representations were added, in an attempt to localize from the 
corresponding massive spectrum the required GUT Higgses. In this way,
extra chiral matter was introduced that survived massless to low energies.
} in the sector $\{ a,
{\tilde O6} \} $. The latter sector, involves open strings
stretching between a D6-brane belonging to the U(5)-stack of
branes and the $\Theta^k$ images of the 06 planes.

\subsection{ { Electroweak Higgses for flipped SU(5)}}

The presence of EW Higgses for flipped SU(5) in the intersecting brane world
context, as in the EW Higgses of the SU(5) GUTS
considered in section (3.1), is independent of the particular lattices on 
which the
tadpole conditions are defined. In fact, their presence is valid for
the tadpole conditions defined in the presence of the AAA, AAB, ABB, BBB,
torus lattices \cite{lust3}, in eqn. (\ref{tad}).

The EW Higgses are `made of' scalar particles \footnote{These particles were not found in \cite{nano} as extra chiral fermions in the representations
$5_2$, ${\bar 5}_{-2}$ were added, in order to localize part of their massive
superpartners - in the same representations - as EW Higgses. In this way,
extra chiral matter was introduced into the 3G chiral content of flipped 
SU(5) that survived massless to low energies.} 
that are part of the
massive spectrum localized in the intersection $\{51^{\star}\}$.
Under the flipped $U(1)^{fl}$ symmetry, the charges of these EW
Higgses localized in the intersection $\{51^{\star}\}$ may be seen
in table (\ref{Higgs1}).
\begin{table} [htb] \footnotesize
\renewcommand{\arraystretch}{1}
\begin{center}
\begin{tabular}{||c|c||c|c|c|c||}
\hline
\hline
Intersection & EW Higgs & repr. & $Q_a$ & $Q_b$ & $Q^{fl}$\\
\hline\hline
$\{ 51^{\star} \}$ & $h_3$  &  ${\bf 5}$   & $1$ & $1$ & $-2$\\
\hline
$\{ 51^{\star} \}$  & $h_4$  & $ {\bf {\bar 5}}$   & $-1$ & $-1$ & $+2$\\
\hline
\hline
\end{tabular}
\end{center}
\caption{\small Higgs fields responsible for EW symmetry breaking in
the flipped $SU(5)$ GUTS.
\label{Higgs1}}
\end{table}
By looking at table (\ref{Higgs1}) we can now argue that the charges of the
Higgs ${\bf 5}$-plets $h_3$, $h_4$ under the flipped U(1)
symmetry, are exactly the one's to play the role of electroweak scalars in a
 flipped SU(5) GUT. 
The same
Higgses are responsible for EW symmetry breaking in the SU(5) GUT of
section three. We must keep in mind that in this case the flipped $U(1)^{fl}$ symmetry was
broken. As it is obvious from table (\ref{Higgs1}), the representation
under which the EW $h_3$ transforms is exactly the conjugate of that of
$h_4$. For clarity of notation we avoided the identification
${\bf 5}$-plet for $h_3$ and its conjugate for
 $h_4$, since their charges are equal and opposite under the flipped
 $U(1)^{fl}$ symmetry.
By repeating the analysis of section (3.1), it is the effective
combination
\beq
h_3 + h_4^{\star}
\eeq
which gets localized at the intersection $\{ 51^{\star} \}$.
Hence, we have shown that the necessary EW Higgses are present
in the flipped classes of GUTS and thus the models will achieve
naturally electroweak symmetry
breaking to $SU(3)_C \otimes U(1)_{EM}$.

\section{Quark-lepton-neutrino masses
in Flipped SU(5)}



\subsection{ Quark masses }


The classes of models we discuss in this work are based on the $Z_3$ orientifold 
backgrounds of \cite{lust3}. We can now make use of the fact that that
the complex structure modulus is fixed at the $Z_3$ orientifold 
backgrounds, in order to discuss the general form of the trilinear Yukawa 
couplings.
It arises
from the stretching of the worldsheet between the D6-branes that
cross at these intersections. Namely two fermions $F_L^i$, $F_R^j$
and one Higgs $H^k$, take the
form, to leading order 
\beq Y^{ijk} \ = \
e^{-A} \ = \ e^{- \frac{R_1 R_2}{\alpha^{\prime}} \ A_{ijk} } \ =
\ e^{- {\sqrt{3} \ R_2^2} \ A_{ijk} } 
\label{form}
\eeq where $A_{ijk}$ may be
of order one.
We assume that the worldsheet areas
involved in the interactions across the two dimensional tori in
the second and third complex planes are close to zero, and thus the 
leading area
dependence arises from the first complex plane.

 In the flipped SU(5) GUTS the u-quark gets naturally a mass
of order $\langle \upsilon \rangle$. Its precise value is
controlled by the $ \Psi^u$ Yukawa coupling
\beq
 \Psi^u \cdot { {\bf 10}} \cdot {\bar {\bf 5}} \cdot {\bar {\bf 5}}^B
\label{exa1}
\eeq
The physical value 
\beq
m_u = \Psi^u \cdot \upsilon \sim e^{-A^d} \cdot \upsilon
\eeq
for the u-quark mass, 
$m_u = 0.005$ GeV is approached hierarchically, from above, by 
having a worldsheet area $A^u = 10.08 $. It is thus naturally 
connected to the electroweak scale of the theory. By implementing relation
(\ref{form}) we find that the value of the u-quark mass is compatible with
a tori radii, $R_2 = 2.4$.
              We note that on hierarchy grounds
$\upsilon$ is the natural scale of the u-quark mass.
We also note the absence of the tree level coupling 
${\bf 10} \cdot  {\bf 10} \cdot{\bf 5}$ normally responsible for the d-quark mass 
generation.
Hopefully, such  a coupling may instead be induced by 
non-renormalizable terms.

\subsection{ Neutrino masses }

The flipped SU(5)  naturally incorporates in its representation context the right handed neutrino
$\nu^c$ neutrino as part of the 10-plet. The see-saw mechanism appears to be
the most plausible
mechanism for generating large values for its mass.

The see-saw mechanism is generated by the interaction \beqa {\cal
L} = {\tilde Y}^{\nu_L \nu_R} \cdot {\bf 10} \cdot {\bar {\bf
5}} \cdot {\bar h}_4 \
      + \  {\tilde Y}^{\nu_R} \cdot \frac{1}{M_s} \cdot ({\bf 10} \cdot
{\bf \overline{10}}^B)
      ({\bf 10} \cdot {\bf \overline{10}}^B),
\label{seesaw1}
\eeqa
where its standard version (\ref{see}) can be generated by choosing
\beq
\langle h_4 \rangle = \upsilon, \  \langle {\bf 10}_i^B \rangle = M_s
\label{masse1}
\eeq

\subsection{Lepton masses }

The mass term for charged leptons, of order of the electroweak scale (times the usual exponential suppression from the Yukawa couplings), exists
naturally within the context of D-brane model building from
the present backgrounds.  It is given by
\beq
Y^{l} \cdot f \cdot l^c \cdot \langle h_3 \rangle + c.c
\eeq
and the hierarchy of the lepton masses is naturally generated by varying
the worldsheet area within the $Y^{l}$ Yukawa coupling.

\section{Proton decay and doublet-triplet splitting}

The embedding of the standard model $SU(3) \times SU(2) \times U(1)$ 
into a higher group, the grand unified group, GUT, was the original idea
\footnote{See  \cite{pati1} for the first realization of this idea
and also \cite{pati2} for a review of problems in gauge theory GUTS including
proton decay.} of
\cite{pati1}. In this context, one has to justify the mass scale of the extra,
beyond the SM gauge bosons involved in the theory. In the present
SU(5) and flipped SU(5) GUT models respectively one has to justify - from 
first principles - the
difference
between the high energy scale $M_{GUT}$, set out by the presence of the extra
gauge bosons mediating the proton decay dimension six
\footnote{where $\epsilon^{ijk}$, $\epsilon^{\alpha \beta \gamma}$,
the totally antisymmetric SU(3) tensor with $\epsilon^{123} \equiv +1$.}
baryon violating \cite{wein} couplings
\beq
p \rightarrow e^{+} \pi^{0}  : \ \sim \frac{1}{M_s^2}\ ({\bar u}^c_L)^i  (u_L)^j \ ({\bar e}_{L, R}^{+}) (d_{L, R})^k \epsilon^{ijk},
\label{proton}
\eeq
\beq
p \rightarrow {\bar \nu}_L \ \pi^{+}  \sim \frac{1}{M_s^2}\
(({\bar d}^c_R)_{\alpha} (u_R)_{\beta}) (({\bar d}^c_L)_{\gamma} \nu_L)\epsilon^{\alpha \beta \gamma}, 
\eeq
\beq
n \rightarrow e^{+} \pi^{-} \sim \frac{1}{M_s^2}\ ( ({\bar u}^c_L)_{\alpha} 
(d_L)_{\beta} )(
({\bar u}^c_R)_{\gamma} e_R ) \epsilon^{\alpha \beta \gamma},
\label{proton1}
\eeq
and the low scale set out by the masses of the quarks and leptons.

\begin{figure}
\centering
\epsfxsize=2.5in
\hspace*{0in}\vspace*{.0in}
\epsffile{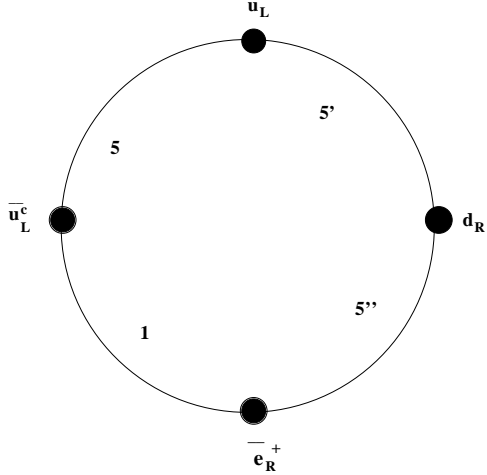}
\caption{\small Proton decay mode $p \rightarrow e^{+} \pi^{0}$ for 
flipped SU(5) GUTS from intersecting branes.
}
\end{figure}
In the present (SU(5) and flipped SU(5)) GUTS, baryon number is not a 
gauged symmetry.
In fact, one can show that gauge boson mediated dimension six
operators, written in an $SU(3) \times SU(2) \times U(1)_Y$ invariant form,
and described by the four-fermion interactions of (\ref{proton} - \ref{proton1})
do exist.
The following proton decay modes are allowed \footnote{In the following 
for simplicity reasons we will omit SU(3) indices.} in SU(5)
\beqa
 & \sim& \frac{1}{M_s^2}\ ({\bar u}^c_L \ u_L) \ ({\bar e}_{L, R}^{+}) 
(d_{L, R}),\\
&\sim &\frac{1}{M_s^2}\ ({\bar d}^c_R \ u_R) ({\bar d}^c_L \ \nu_L),
\\
&\sim &\frac{1}{M_s^2}\ ( {\bar u}^c_L \ d_L  )
({\bar u}^c_R \ e_R )
\eeqa
and flipped SU(5) GUTS
\beqa
 & \sim& \frac{1}{M_s^2}\ ({\bar u}^c_L \ u_L) \ ({\bar e}_{R}^{+}) 
(d_{R}),\\
&\sim &\frac{1}{M_s^2}\ ({\bar d}^c_R \ u_R) ({\bar d}^c_L \ \nu_L) \ .
\eeqa
The relevant disc diagrams can be seen in figures 1-5. The numbers shown in
the boundary indicate the $D6_a$ brane that the disk boundary is attached.

\begin{figure}
\centering
\epsfxsize=2.5in
\hspace*{0in}\vspace*{.0in}
\epsffile{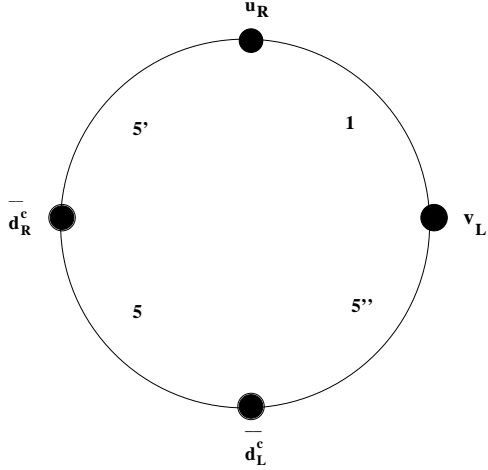}
\caption{\small Proton decay mode $ p \rightarrow {\bar \nu}_L \ \pi^{+}$ 
for flipped SU(5) GUTS from intersecting branes.
}
\end{figure}

Thus for example the depicted proton decay mode associated with $e_L^{+}$ 
in figure 3 
is associated with the 
SU(5) structure ${\bf 10}^2 \cdot {\bf {\overline{10}}^2}$ 
and has been calculated in a N=1 supersymmetric context 
in \cite{igorwi}, for another class of models. 
On the other hand flipped SU(5) offers another kind of decay modes - seen in figures 1 and 2 -
which have the SU(5) structure  ${\bf 10} \cdot {\bf 5} \cdot 
{\bf {\overline{10}}} \cdot {\bf {\bar 5}}$. 
It will be interesting to perform the calculation of the latter 
amplitude, to see the differences of the flipped SU(5) proton decay 
amplitude against the amplitude studied in \cite{igorwi}.

In addition, we note the presence of the decay 
modes $p \rightarrow e^{+} \pi^{0} $, $ p \rightarrow \pi^{+} \nu$ associated
with the SU(5) structure  
${\bf 10} \cdot {\bf {\overline{10}}}\ {\bf 5} \cdot {\bf {\bar {5}}} $. The presence of these modes which exist in a general
global four dimensional GUT, did not arise in the models examined in
\cite{igorwi}.
We also note that the same amplitudes may exist in a N=1 
supersymmetric context of the models as the charge structure which 
supports these reactions, essentially the chiral matter 
context of the models incorporating the three generation representations, 
remains identical. It will also be interesting to perform these calculations.

As the reactions (\ref{proton} - \ref{proton1}) associated to the various
decay modes, 
get mediated by the extra, beyond the SM gauge bosons
 with a mass \footnote{For simplicity we consider that all the 
beyond the SM gauge bosons have the same mass, namely $M_X$.}
of order $M_X$,  we have to rather
choose $M_X = M_s = 10^{16}$ GeV, 
in order to enhance the proton decay rate beyond experimental
observation. 
Related observations have been made in \cite{igorwi}.

Also we note that for a GUT scale of order of the $10^{16}$ GeV, proton decay modes mediated by 
dimension six operators are the dominant one's.\newline
In this case, gauge mediated proton decay
may be enhanced beyond the superK bound \cite{superK}
$\Gamma^{-1}(p \rightarrow e^{+} \pi^0)_{expected} \geq 10^{33}$ yrs.
However, this is not enough to save the proton. It is expected that there
will also be scalar mediated proton decay modes in a general GUT model.
These dangerous triplet 
scalars are part of the electroweak Higgses for SU(5) GUTS, but
for GUTS based \footnote{A flipped SU(5) GUT is not really
a GUT in the sence of the simple SU(5) group but, in this context, neither 
it is the original Pati-Salam GUT
$SU(4)_c \times SU(2)_L \times SU(2)_R$ \cite{pati1}.}
on the flipped SU(5) GUT \cite{bar}, these scalars
are part of the electroweak 5-plets and GUT 10-plets, and have been
denoted earlier as $D$, $d^c_H$ respectively.

In fact, 
baryon number violating problems associated with the existence 
of a light mass for these triplets may be absent, if we can show
that the coloured triplet mediating scalars have a mass
of order $M_{GUT}$.  This constitutes the famous doublet-triplet splitting (DTS)
problem \footnote{For global SUSY SU(5) GUTS it was discussed in \cite{be}.}.
\begin{figure}
\centering
\epsfxsize=2.8in
\hspace*{0in}\vspace*{.0in}
\epsffile{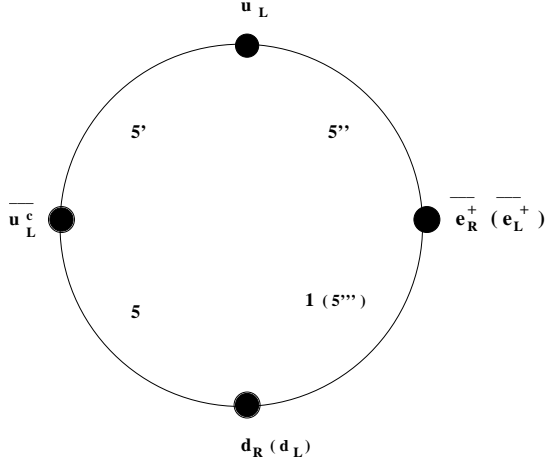}
\caption{\small Proton decay mode $  p \rightarrow e^{+} \pi^{0}$ 
for SU(5) GUTS from intersecting 
branes.
}
\end{figure}
In a N=1 
supersymmetric context, the DTS solution demands the presence of certain mass
couplings that have its origin in the superpotential of the theory. 
This is to be contrasted with the origin of the DTS solution in a 
non-supersymmetric
context, which demands the presence of the appropriate couplings
in the effective scalar potential.  In this case the scalar 
potential is able to break the GUT symmetry and simultaneously give masses to the coloured
 triplets. Thus the usual SU(5) theories, as the one's examined 
in the first part of this work solve the DTS problem directly using
the effective scalar potential \cite{bu}.

Let us now focus our attention to the DTS solution of the flipped SU(5) GUTS.
 In the SUSY SU(5) DTS is achieved
by using large representations that may be absent in string theories.
On the other hand the field theoretical version of N=1 SUSY
flipped SU(5) \cite{GG} offers a most 
economical solution to the gauge hierarchy
problem by using \footnote{The solution of the doublet-triplet problem
proceed along similar lines in the fermionic formulation version of
flipped SU(5) \cite{AB}. } only the 10-plets part of the GUT scalars,
that allow the DTS solution to arise from the presence of the term
$HHh$ in the superpotential of the theory.

We note that in the past, there have been attempts in a N=1 
four dimensional heterotic string compactification context, to solve 
the doublet-triplet 
splitting by excluding the presence of triplet scalars with either
the use of 
continuous Wilson lines \cite{ibaba, fara}, or by the use of 
discrete symmetries in the context of M-theory 
compactifications \cite{wit}.

\begin{figure}
\centering
\epsfxsize=2.9in
\hspace*{0in}\vspace*{.0in}
\epsffile{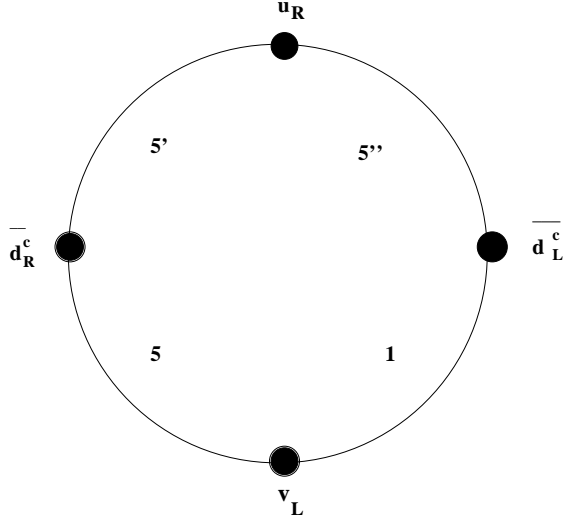}
\caption{\small Proton decay mode $  p \rightarrow {\bar \nu}_L \ \pi^{+}$ for 
SU(5) GUTS from intersecting 
branes.
}
\end{figure}

 Because, from charge conservation, the term $HHh$ is not allowed in the present
flipped SU(5) GUTS the invariants that may be written by using 
this term involve a variety of physical fields.
Their form 
necessarily will be based on higher non-renormalizable terms.
We keep in mind that
our GUT classes of models are non-supersymmetric and thus a direct use of 
a superpotential tool is not possible. Instead, we may write down the
solution to the doublet-triplet splitting problem by writing down the
following 
effective scalar potential terms 
\beqa
\frac{r}{M_s^3}(HHh)( {\bar F}  {\bar F} {\bar h}) + m ({\bar h}h) (
{\bar H} H) + \kappa ({\bar H}H)( 
{\bar H} H),\\
\frac{r}{M_s^3}(\langle H \rangle Hh)({\bar F}{\bar F} \langle {\bar h} \rangle) + m ({\bar h}h) (\langle 
{\bar H}\rangle \langle H\rangle) + \kappa ({\bar H}H)(\langle 
{\bar H}\rangle \langle H\rangle)
\label{yuk1}
\eeqa
\beq
r\frac{M_{GUT} \cdot \upsilon \cdot \langle d_L d_R \rangle  }{M_s^3}(d_c^H D) + m \cdot(D{\bar D})(  \langle H \rangle  
\langle H \rangle  
+ \kappa\cdot{\bar d^c_H}{d^c_H} ( \langle {\bar H}\rangle \langle  H \rangle)
\label{effe}
\eeq
where $r, m, k$ Yukawa coupling coefficients. From
(\ref{effe}) we get the eigenvalues
\beq
M_{D} \approx m^{1/2} M_{GUT}\ (1 + \frac{r^2}{(m-k)m} \omega), \
M_{d^c_H} \approx \kappa^{1/2} M_{GUT}(1 -\frac{r^2}{(m-k)k} \omega),
\eeq
for the $d^c$, $D$ triplets respectively, where we have defined

\beq
\lambda =  \frac{M_{GUT} \cdot \upsilon \cdot \langle d_L d_R \rangle}
{M_s^3}, \  
\omega =  \frac{\upsilon^2 \langle d_L d_R 
\rangle^2 }{ M_{GUT}^2  M_s^6}
\eeq
\beq
m^{1/2} \sim  e^{-\frac{A_m}{2}}, \  
k^{1/2} \sim  e^{-\frac{A_k}{2}}, 
\eeq
where $A_m$, $A_k$ the worldsheet areas associated with the 
couplings seen in (\ref{yuk1}).
Note that because of the presence of the triplet mixing term 
in (\ref{yuk1}) we don't have to choose the GUT scale equal to the 
string scale. \newline The triplet mixing term affects only slightly 
the triplet masses and its presence can test the fluctuation of the values of
the triplet masses around the GUT scale. In fact, the invariants involved in 
the triplet mass matrix (\ref{effe}) are the lowest dimensional one's 
consistent with gauge invariance and charge conservation. 

\begin{figure}
\centering
\epsfxsize=2.8in
\hspace*{0in}\vspace*{.0in}
\epsffile{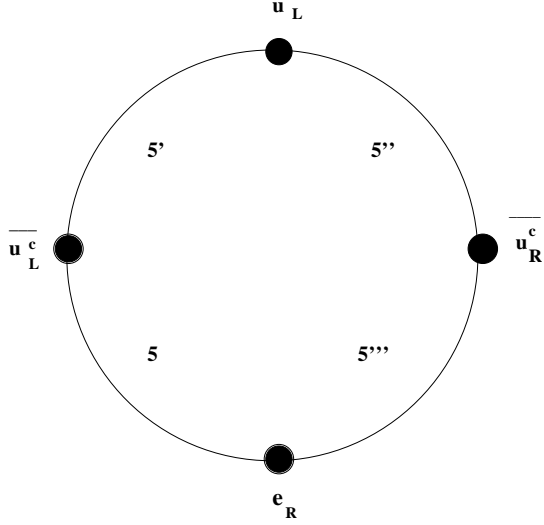}
\caption{\small Proton decay mode $  n \rightarrow e^{+} \pi^{-}    $ for SU(5) GUTS from intersecting 
branes.
}
\end{figure}

By choosing $M_{GUT} = 10^{16}$ GeV,
$M_s = 10^{17} GeV$, $\upsilon = 246$ GeV and substituting in the value of the chiral 
condensate $<d_L d_R> \approx (220 MeV)^3$ \cite{vl}, we find that 
\beq
\omega =  {\cal O}(10^{-134})
\label{esti}
\eeq 
which is a very small number. Thus the triplet mixing term leaves for 
all practical purposes unaffected the triplet masses of order of the
GUT scale. Clearly, the same conclusion may be reached if higher order
non-renormalizable contributions to the triplet mass matrix are 
included.

\section{Conclusions}

In this work, we have presented general structural elements of
four dimensional SU(5) GUT intersecting D6-brane model building,
which are included in a large class of type IIA orientifold with
orbifold symmetries \cite{lust3}.

The basic properties enjoyed by these GUT constructions are :

$\bullet$ RR tadpole cancellation, which guarantees the
cancellation of the cubic gauge anomaly in the effective theory of light 
modes.

$\bullet$ absence of NSNS tadpoles associated with the complex structure 
moduli. We note that as a dilaton tadpole remains at one loop level,
the dilaton runs away to zero coupling \footnote{See \cite{lust3} for a discussion how this latter 
problem might be cured. Also, note that running dilaton moduli is a 
general property of N=1
heterotic string compactifications without fluxes.}.

We explored two possibilities in SU(5) GUT model building, 
one with a purely SU(5) symmetry and one with a flipped $SU(5) 
\times U(1)$ symmetry.
In this work we have lifted the restriction of the 
electroweak pentaplet Higgs identification. Also, we have appropriately
identified the GUT breaking scalars for general flipped SU(5) GUTS
(also applicable to $Z_N$ and $Z_N \times Z_M$ orientifolds). Let us now review
the phenomenological properties of the SU(5) GUTS considered 
in the present work.

In both cases, SU(5) and flipped SU(5) type of GUTS respectively :

$\bullet$ we presented a model which does not include any
surviving at low energies exotic matter states, and thus the models have 
only the SM at low energy. They may also achieve naturally the electroweak symmetry 
breaking as a consequence
of the appropriate identification of the electroweak scalar
pentaplet Higgses.

$\bullet$ We showed that the phenomenological features of these GUTS,
include hierarchical values (HV) for neutrino masses (eqn's \ref{seesaw}, 
\ref{seesaw1} respectively) due to the
see-saw mechanism \cite{kokos1}, and also tree level HV Yukawa couplings for leptons 
and 
for the $(d, s, b)$, $(u, c, t)$ quarks respectively. We observe that we 
have not
been able, to generate higher order non-renormalizable Yukawa
couplings for the $(u, c, t)$, $(d,s, b)$ quarks respectively.
This is due to the fact that the usual tree level coupling term $10 \cdot 10 \cdot 5 $ coupling which is of order of the electroweak
symmetry breaking scale $\upsilon$,
is not allowed by charge conservation. The resolution of this problem may be 
attibuted to non-norenormalizable couplings. We hope to return to this issue
in the future.

$\bullet$ We note that while baryon number is not a gauged symmetry in the 
SU(5) and flipped SU(5) GUTS constructed in this work from intersecting brane 
worlds, gauge mediated
dimension six operators do exist. They may be appropriately getting 
suppressed if a large GUT scale is chosen of the order of the $10^{16}$ 
GeV. Moreover, the solution to the doublet-triplet splitting problem
guarantees the suppression of scalar mediated proton decay dimension six
operators.  
In conclusion, non-supersymmetric models
from intersecting brane world orbifolds may be fully safe against proton
decay. 

Finally, because the scale of the models is large enough 
one expects quadratic loop corrections to the EW Higgs masses to 
appear. This, in fact, we identify
as the low energy manifestation of the gauge hierarchy problem.
Indeed as the models may possess N=1 supersymmetric sectors
 for particular choices 
of wrapping numbers,  one might expect that quadratic Higgs divergences
may cancel at one loop \cite{iba} (and possibly higher). Thus a full solution to the 
gauge hierarchy at the weak scale remains an open issue in the present classes
of non-supersymmetric GUTS.

\begin{center}
{\bf Acknowledgments}
\end{center}
We are grateful
to I. Antoniadis, L. Ib\'a\~nez, S. Sint, and A. Uranga,
for useful discussions. C.K would like to thank the Institute of
Nuclear Physics Dimokritos for hospitality and support during the course of
this work and also the organizers of
the conference SUSY 2003,
for their warm hospitality,
 where part of this work was done.

\end{document}